\documentclass[aps,reprint,twocolumn,showpacs,superscriptaddress]{revtex4-1}

\usepackage{graphicx}
\usepackage{dcolumn}
\usepackage{bm}
\usepackage{amsfonts}
\usepackage{amsmath}
\usepackage{amssymb}
\usepackage{color}
\usepackage[normalem]{ulem}

\usepackage{xcolor, soul}
\sethlcolor{green}
\usepackage{mdframed}
\usepackage{mathrsfs}   
\usepackage[percent]{overpic}

\usepackage[colorlinks=true,citecolor=blue]{hyperref}
\hypersetup{colorlinks=true,citecolor=blue,linkcolor=red,urlcolor=blue}

\def\be{\begin{equation}}
\def\ee{\end{equation}}

\def\kv{{\bf k}}
\def\qv{{\bf q}}

\newcommand{\h}[1]{{\hat {#1}}}
\newcommand{\hdg}[1]{{\hat {#1}^\dagger}}

\begin{document}

\title{Phase-diagram and dynamics of Rydberg-dressed fermions in two-dimensions}

\author{Reyhaneh Khasseh}
\affiliation{Department of Physics, Institute for Advanced Studies in Basic Sciences (IASBS), Zanjan 45137-66731, Iran }
\author{Saeed H. Abedinpour}
\email{abedinpour@iasbs.ac.ir}
\affiliation{Department of Physics, Institute for Advanced Studies in Basic Sciences (IASBS), Zanjan 45137-66731, Iran }
\author{B. Tanatar}
\affiliation{Department of Physics, Bilkent University, 06800 Ankara, Turkey}

\date{today}

\begin{abstract}
We investigate the ground-state properties and the collective modes of a two-dimensional two-component Rydberg-dressed Fermi liquid in the dipole-blockade regime. We find instability of the homogeneous system toward phase separated and density ordered phases, using the Hartree-Fock and random-phase approximations, respectively. The spectral weight of collective density oscillations in the homogenous phase also signals the emergence of density-wave instability. We examine the effect of exchange-hole on the density-wave instability and on the collective mode dispersion using the Hubbard 
local-field factor.
\end{abstract}

\maketitle

\section{Introduction}\label{sec:introduction}
Studying ultracold Rydberg systems is becoming a fascinating field from both fundamental and practical points of view.
Such artificial systems provide a mean to explore many-body models in a very controllable manner~\cite{controlable_Reyd}.  
On the applied side, these systems are promising for the quantum manipulations and computations~\cite{quantum_inform}.
Different experimental methods, such as laser cooling and magnetic trapping~\cite{laser_cool,laser_cool2} of neutral atoms, have recently been employed.
In order to reach nano-Kelvin temperature regime, the magneto-optical trapping is an appropriate technique~\cite{mot}.
The traditional example of Rydberg atoms is hydrogen. However, any alkali metal element can be excited to a higher energy level, with a very high principal quantum number~$n$, in which the valence electron obeys well-known Rydberg energy spectrum~\cite{hydrogen_like}. The effective size of these atoms can be very large, up to the micrometers~\cite{dim_Reyd}. 
\par
Although the total charge of Rydberg atoms is zero, they can have very strong dipole moments. In this regard, there is a strong dipole-dipole interaction among Rydberg atoms. One can take advantage of this long-range interaction in order to construct strongly correlated systems. Exploring these systems seems to be very appealing and one may expect to observe some new and exotic quantum phenomena and behavior. Notice that the interaction between Rydberg atoms is very similar to the van der Waals interaction $\propto 1/r^6$, at large enough inter-particle distances~\cite{Riv_Reyd, intraction_Reyd, RF_book, dipole_strength}. However, for short inter-particle distances, called the {\it Rydberg-blockade radius}, the strong dipole-dipole interaction affects the binding energies of valence electrons in each atom and make them off-resonance with respect to the laser frequency. As a result, the Rydberg energy spectrum is destroyed and the $1/r^{6}$ interaction evolves into a soft-core repulsive interaction with finite interaction range. The typical value of the dipole blockade radius, $r_c$, is around $5-10~{\rm \mu m}$~\cite{block_Reyd,block2_Reyd,Reyd_Bolkade,Reyd_block}. 
In order to construct a Rydberg atom with a long lifetime, two laser beams where only the sum of their frequencies is on-resonance with the Rydberg levels can be used. In this way, a Rydberg atom through a two-photon absorption process will be made. The resulting atom is called {\it Rydberg-dressed} atom which has a longer lifetime with respect to the conventional Rydberg atoms made through single-photon absorption~\cite{metallic_quantum_phase, Reyd_dreesed, Reyd_intraction}.
\par
The effect of dipole-dipole interaction results in some interesting physics. For example, 
density-wave instability in a three-dimensional (3D) system of Rydberg-dressed fermionic atoms, interacting through the van der Waals force, is predicted to lead to a body-centered-cubic (BCC) crystalline ordering with gapless fermionic excitations, named \emph{metallic quantum solid phase}~\cite{metallic_quantum_phase}. 
The emergence of superfluid vortices in a rotating two-dimensional (2D) Bose-Einstein condensate (BEC) of Rydberg-dressed atoms has been studied in Ref~\cite{supersolid_Reyd}.
A transition to a Bardeen-Cooper-Schrieffer (BCS) state has been predicted in a system of Rydberg-dressed atoms under repulsive van der Waals interaction, assisted by BEC of diatomic molecules which play the analogous role of phonons in the BCS superconductivity~\cite{Ry_BCS}.
\par      
Naturally, it is expected that the effects of many-body correlations would be enhanced at lower dimensions. Furthermore, reduced dimensionality is expected to suppress collisions and chemical reactions~\cite{baranov_chemrev2012}.
Motivated by these, in this paper we investigate the ground-state phases of a two-component 2D system of Rydberg-dressed atoms. 
Using the Hartree-Fock (HF) mean field approach, we show that a quantum phase transition equivalent to the ferromagnetic-paramagnetic phase transition in an electron liquid is possible in two-component Rydberg-dressed fermionic liquid (RDL), by tuning the average density and Rydberg-blockade radius.
The term ``two-component" here, refers to two different fermionic isotopes of the same atom, or identical atoms in two different internal states. Therefore, the above-mentioned phase transition corresponds to the phase transition in the first case and to the internal degree polarization in the second case.  For brevity, in the following, we will simply refer to this phase transition as \emph{phase separation}. 
Notice that the phase separation state has been also predicted for fermionic systems with short-range s-wave scattering~\cite{swave_phase_diagram}.
Moreover, we investigate the density-wave instability (DWI) of a homogenous phase-separated (\emph{i.e.}, single-component) RDL from the singularities of its static density-density response function within the random-phase approximation (RPA). 
Furthermore, we find analytical results for the dispersion of collective density (\emph{i.e.}, zero-sound) modes and the sound velocity in the phase-separated state, using RPA. 
We also improve upon the RPA, including the effects of exchange-hole through the so-called Hubbard local-field factor (LFF)~\cite{GV_book}, and investigate its effects on the DWI and on the dispersion of collective mode.
\par
The rest of this paper is organized as follows. 
In Sec.~\ref{sec:formalism} we describe our model Hamiltonian and the effective interaction between Rydberg-dressed particles. 
In Sec.~\ref{sec:HF} we employ the Hartree-Fock mean-field approximation to calculate the ground-state energy of a two-component Rydberg-dressed system and investigate the possible phase transition between homogenous mixed and phase-separated states. In Sec.~\ref{sec:DWI}, we calculate the static dielectric function of RDL within the RPA and study the instability of the system towards the formation of density-waves.
We study the zero-sound mode of a homogeneous RDL using RPA in Sec.~\ref{sec:ZSM}. 
Finally, we summarize our main results and conclude in Sec.~\ref{sec:conclusion}.
\section{Model and Hamiltonian}\label{sec:formalism}
We consider a 2D system of two-component Rydberg-dressed fermionic particles, where each particle is weakly coupled
to its Rydberg state by an off-resonant two-photon transition. The Hamiltonian of this system reads
\be\label{eq:hamil}
\begin{split}
{\cal H}&=\sum_{\kv, \sigma}\varepsilon_{k,\sigma} \hdg{c}_{\kv,\sigma}  \h{c}_{\kv,\sigma} \\
&+\frac{1}{2S}\sum_{\qv} v_{\rm R}(q)
  \sum_{\kv,\sigma}\sum_{\kv',\sigma'} 
   \hat c^{\dagger}_{\kv-\qv,\sigma} 
  \hat c^{\dagger}_{\kv'+\qv, \sigma'} \hat c_{\kv',\sigma'} \hat c_{\kv, \sigma}~,
\end{split}
\ee
where $\varepsilon_{k,\sigma}= \hbar^2 k^2/(2m_\sigma)$ is the non-interacting energy dispersion of particles of mass $m_\sigma$, $\sigma=A,B$ refers to two different components of the Rydberg particles, $\hat c_{\kv,\sigma}$ ($\hat c^{\dagger}_{\kv,\sigma}$) destroys (creates) one Rydberg-dressed atom of type $\sigma$ with wave vector $\kv$, $S$ is the sample area, and $v_{\rm R}(q)$ is the Fourier transform of the interaction between two Rydberg dressed atoms, which for simplicity we assume not to depend on the types of two atoms $\sigma$ and $\sigma'$. Moreover, we will assume that the mass difference between the two-components is negligible and we set $m_{\rm A}=m_{\rm B}=m$ in all our subsequent analysis. 
\par
\begin{figure}
 \includegraphics[width=\linewidth]{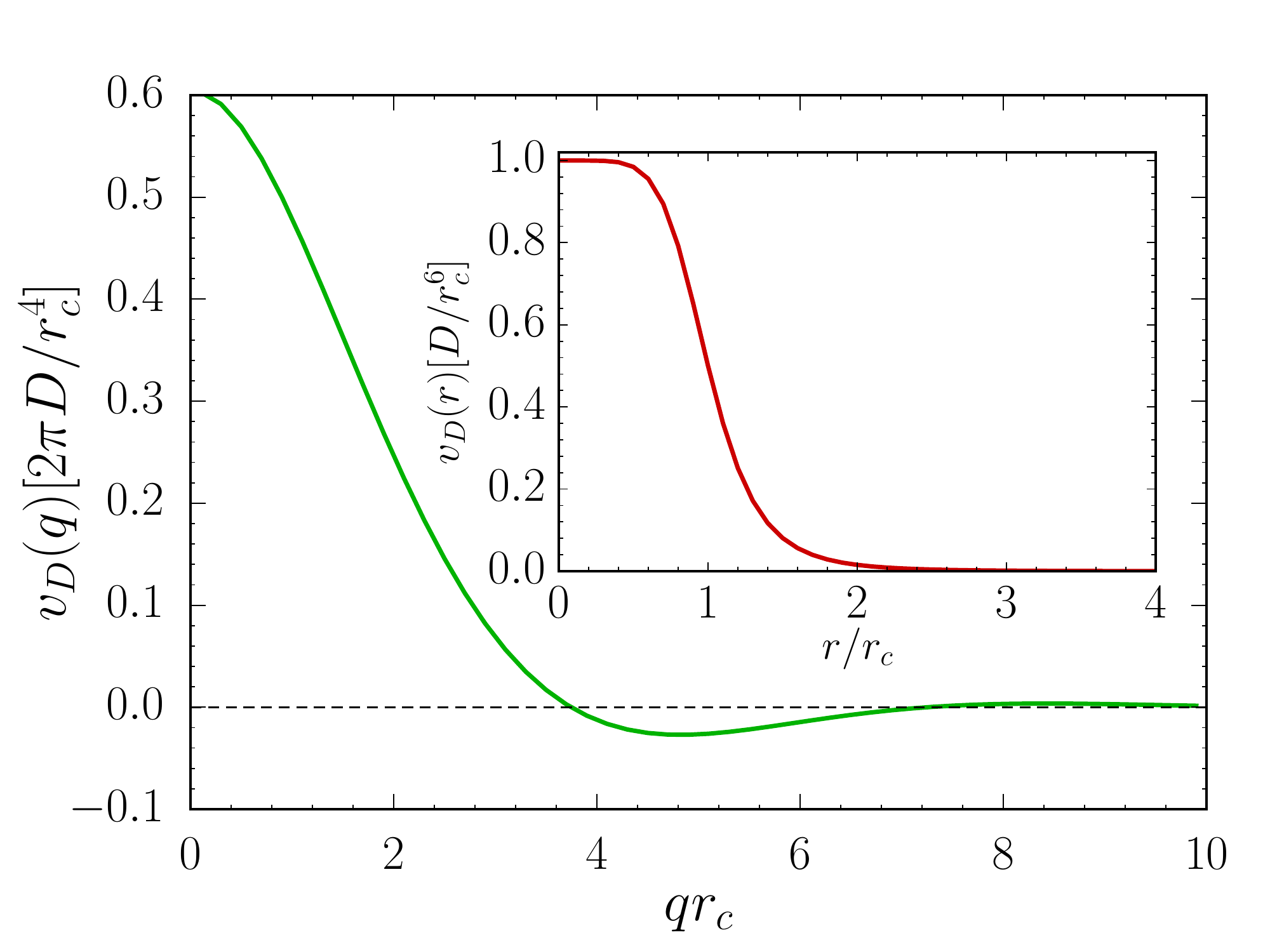}
\caption{Effective Rydberg-dressed potential in momentum space $v_{\rm R}(q)$ (in units of $2\pi D/r_c^4$) versus $q \,r_c$. The inset shows the real space potential $v_{\rm R}(r)$ (in units of $D/r_c^6$) versus $r/r_c$. \label{fig:Ry_potential}}
\end{figure}

The repulsive interaction between two Rydberg atoms is dominated by the van der Waals form when the inter-particle distance is large enough. For smaller distances, close to the Rydberg-blockade radius, this interaction becomes soft. Therefore, we can approximate the inter-particle interaction as~\cite{Red_Wan_interaction,Wan_block}
\begin{equation}
v_{\rm R}(r)=\frac{D}{r^{6}+r_{c}^{6}}~,     
\end{equation}
where $r$ is the 2D distance between two atoms, $D$ is the positive van der Waals coefficient, and $r_c$ is the dipole-blockade radius given by~\cite{intraction_Reyd,Ry_BCS}
\begin{equation}\label{eq4}
    r_{c}=(\frac{2 D \delta}{\hbar \Omega^2})^{1/6}~.
\end{equation}
Here $\Omega$ and $\delta$ are the single-atom Rabi frequency and the detuning, respectively. 
The Fourier transform of $v_{\rm R}(r)$ is obtained as
\begin{equation} \label{eq5}
  v_{\rm R}( q)=\frac{\pi D}{3r_{c}^{4}} G^{4,0}_{0,6}
  \left(\frac{r_{c}^6 q^6}{6^6}\left|_{0,\frac{1}{3},\frac{2}{3},\frac{2}{3},0,\frac{1}{3}}\right.\right)~,
\end{equation}
in terms of the Meijer-G function~\cite{Gradstein_Ryzhik, wolfram} $G^{ij}_{mn}(\dots)$. In Fig.~\ref{fig:Ry_potential} the behavior of inter-particle interaction potential is illustrated in both position and momentum spaces. As can be seen in this figure the interaction potential in the position space is almost flat for $r < r_c$ and has a Van der Waals tail at $r \gg r_c$. Although this interaction is purely repulsive in the position space, its Fourier transform $v_{\rm R}(q)$, has a negative minimum at $q\approx 5/r_c$. This peculiar behavior arises from the dipole-blockade effect and is not observed in other kinds of long-range interactions~\cite{metallic_quantum_phase}.   
%
\par
Before turning to report our results, we should note that the zero temperature properties of a 2D single-component fermionic RDL is characterized by two parameters, the dimensionless density parameter $\lambda = k_{\rm F} r_0$, and the dimensionless blockade radius ${\tilde r}=r_c/r_0$, where $k_{\rm F}=\sqrt{4\pi n}$ is the Fermi wave-vector of a single-component system, $n$ being its particle density, and
$r_0=\left({m D}/{\hbar^2}\right)^{1/4}$~.
A two-component system requires one extra parameter specifying the density imbalance $p=(n_{\rm A}-n_{\rm B})/(n_{\rm A}+n_{\rm B})$, where $n_\sigma$ is the particle density of $\sigma$-component. The density imbalance varies between $p=0$ for a two-component system with equal population of both ingredients, to $p=1$ for a single-component or fully polarized system (note that we name the higher density component as $n_{\rm A}$).      
\par
\section{Ground-state energy and phase separation}\label{sec:HF}
\begin{figure}
 \centering
 \begin{overpic}[width=\linewidth]{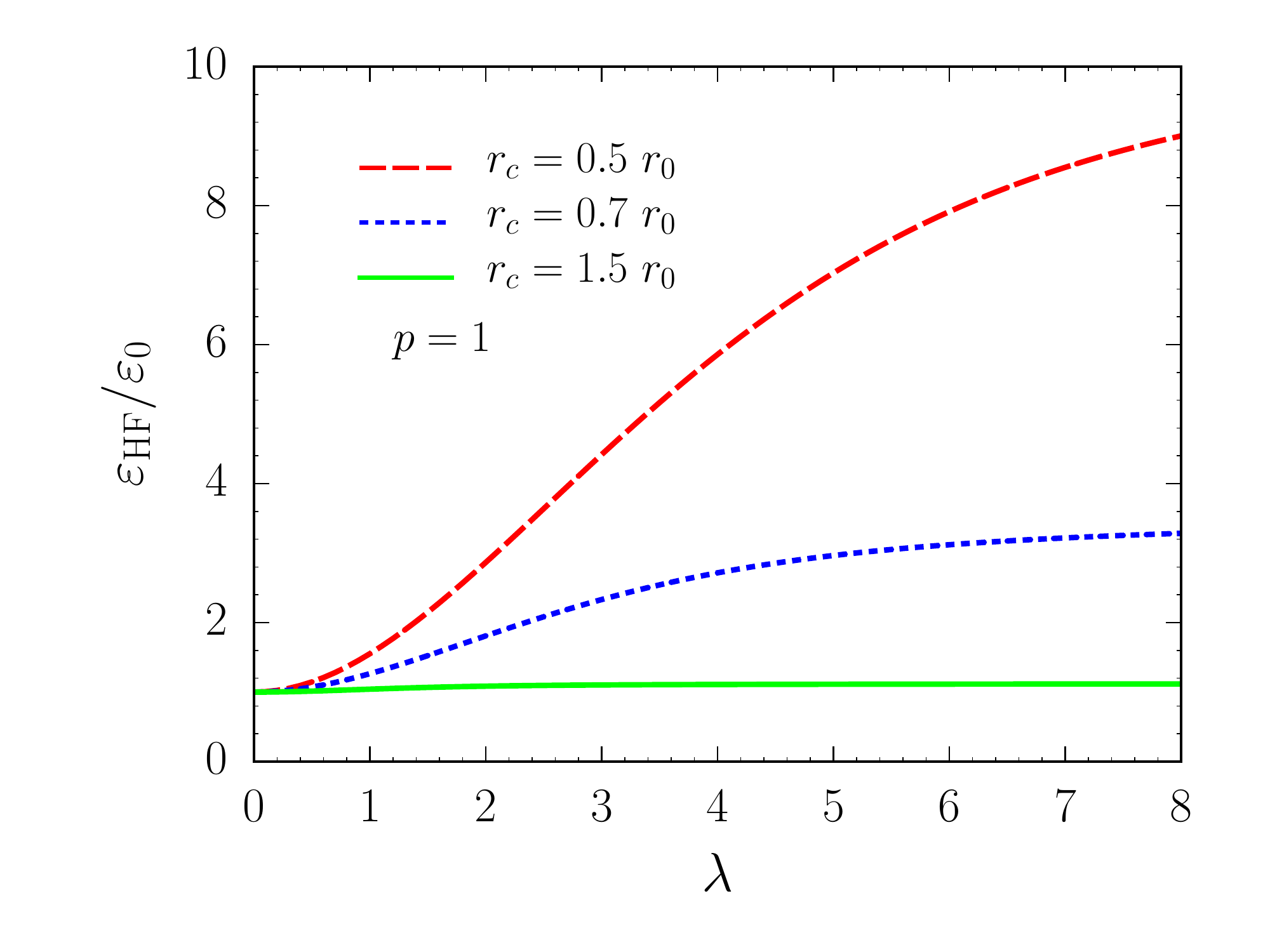}\put(-1,71){(a)}\end{overpic}
 \begin{overpic}[width=\linewidth]{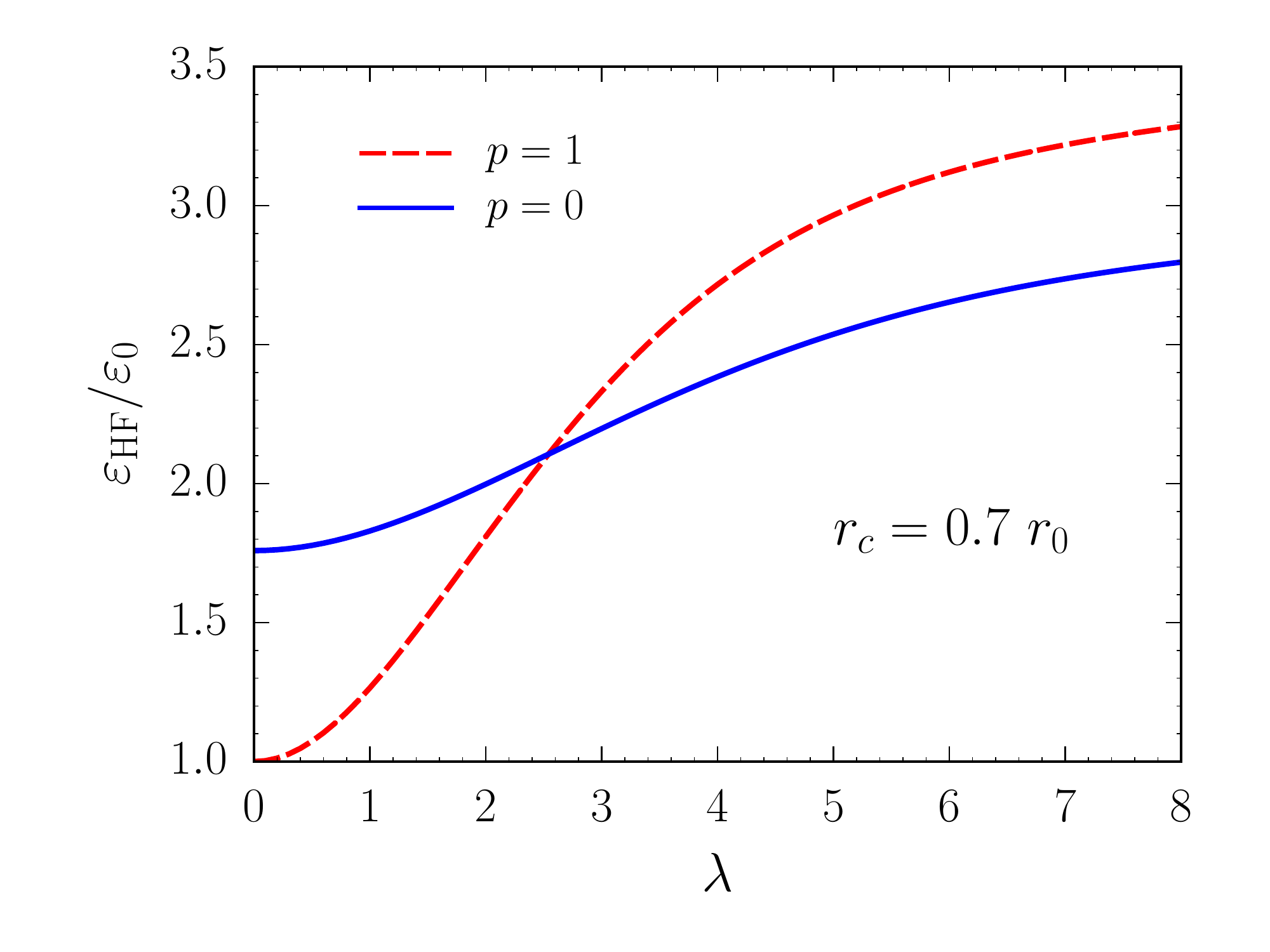}\put(-1,71){(b)}\end{overpic}
 \caption{(a) The ground-state energy of a single-component 2D Rydberg-dressed liquid within the HF approximation (in units of the noninteracting ground-state energy $\varepsilon_0=E_0\lambda^2/4$ of a single-component system), versus the coupling strength $\lambda$ for several values of the dimensionless blockade radius $r_c/r_0$.
  (b) Comparison of the HF energies of mixed and phase-separated states (in units of $E_0\lambda^2/4$) versus $\lambda$ for $r_s=0.7\, r_0$. The phase separated state become lower in energy for $\lambda \lesssim 2.5$. 
  \label{fig:PD}}
\end{figure}

\begin{figure}
 \includegraphics[width=\linewidth]{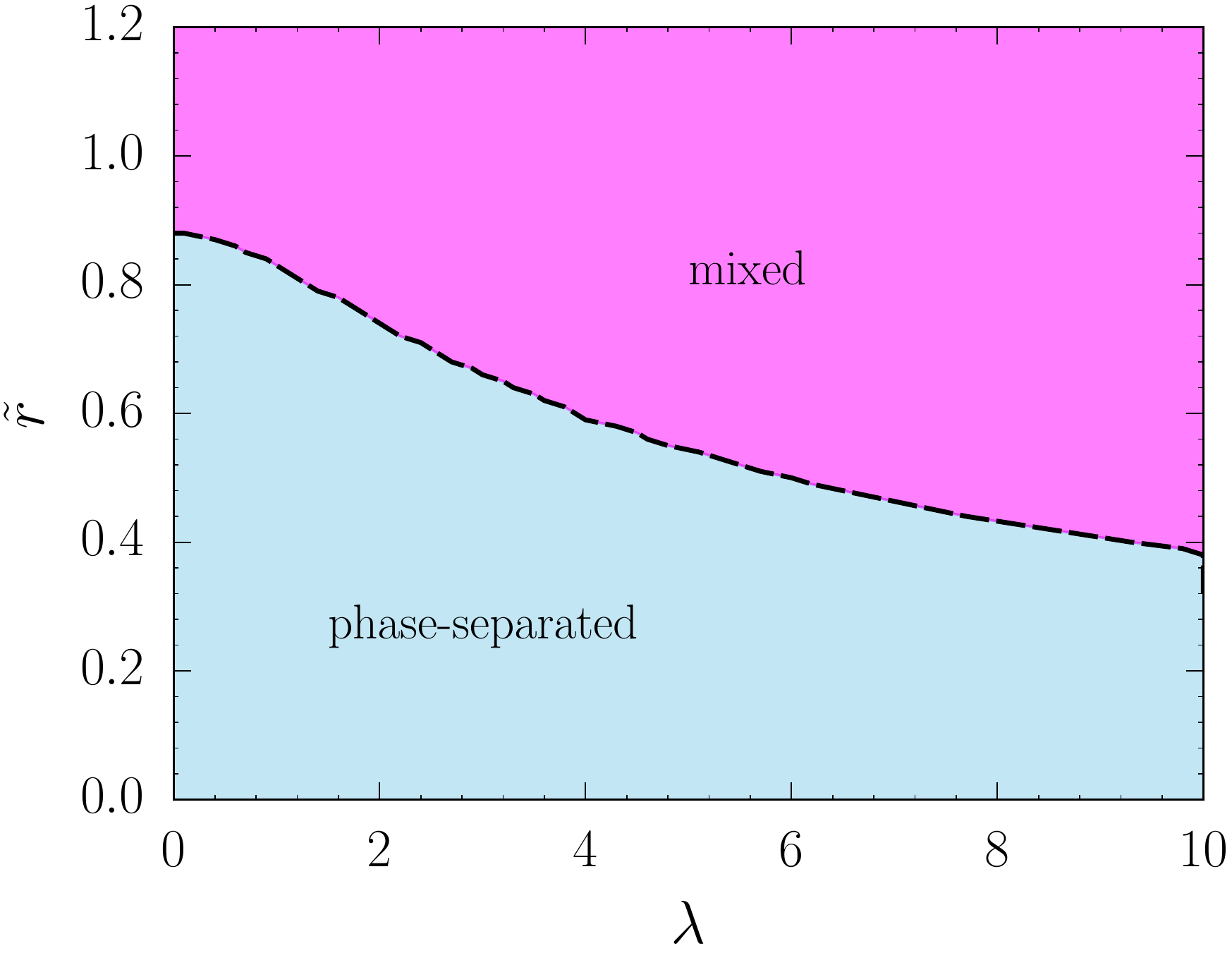}
\caption{The competition between mixed and phase-separated states of a two-dimensional two-component Rydberg-dressed system in the $\lambda$-${\tilde r}$ parameter space, obtained within the Hartree-Fock approximation. \label{fig:mixed-separated}}
\end{figure}

In this section, we first calculate the ground-state energy of a homogenous RDL as a function of $\lambda$, ${\tilde r}$ and $p$, using the HF approximation. Then, we investigate the possibility of transition form a homogeneous mixed phase to a separated one. 
The mean-field energy of two-component system can be evaluated from 
\be\label{eq:HF}
\begin{split}
\langle {\cal H}\rangle_{\rm HF} &=
\sum_{\kv,\sigma} \varepsilon_{k} n_{k,\sigma}+\frac{1}{2A}v_{\rm R}(q= 0) \sum_{\kv,\sigma}n_{k,\sigma} \sum_{\kv',\sigma'}  n_{k' ,\sigma'}  \\
& -\frac{1}{2A}\sum_{\qv} v_{\rm R}(q)  \sum_{\kv, \sigma}  n_{\kv-\qv, \sigma}n_{k ,\sigma}~.
\end{split}
\ee
Here, $n_{k,\sigma}$ is the Fermi occupation function which reduces to a Heaviside step function $n_{k,\sigma}=\Theta(k_{{\rm F},\sigma}-k)$ at zero temperature, where $k_{{\rm F},\sigma}=\sqrt{4 \pi n_{\sigma}}$, is the Fermi wave vector of component $\sigma$.
Note that in the following all the energies will be reported in the units of $E_0=\hbar^2/(m r_0^2)$. 
The dimensionless HF energy \emph{per particle} reads
\be\label{eq:e_hf}
\begin{split}
&\varepsilon_{\rm HF}(p,\lambda,{\tilde r})
=\frac{\lambda^2}{8}\left(1+p^2\right)+\frac{\pi}{12\sqrt{3}}\frac{\lambda^2}{{\tilde r}^4}\\
&-\frac{1}{2 {\tilde r}^6}
\left[1
-\frac{1}{6\sqrt{3\pi}}\sum_{\xi=\pm 1} (1+ \xi p)g\left(\sqrt{1+\xi p}\lambda {\tilde r} \right) \right]
~,
\end{split}
\ee 
where the terms on the right hand side are the noninteracting kinetic, Hartree, and exchange energies, respectively, and
\be
 g(x) = G^{4,3}_{3,9}
        \left[\left(\frac{x}{3 \sqrt{2}}\right)^6\left|^{\frac{1}{6},\frac{1}{2},\frac{5}{6}}_{0,\frac{1}{3},\frac{2}{3},1,-\frac{1}{3},0,\frac{1}{3},\frac{1}{3},\frac{2}{3}}\right.\right]~.
\ee
Interestingly, both the kinetic and Hartree terms have quadratic dependance on the coupling constant $\lambda$. 
In Fig.~\ref{fig:PD}\,(a) the $\lambda$ dependance of the HF energy of a single-component (\textit{i.e.}, $p=1$) 2D RDL has been illustrated. As expected, the effects of direct and exchange interactions are substantially suppressed at large blockade radii and the system essentially becomes noninteracting. At low densities ({\it i.e.}, $\lambda \to 0$), using the fact that
\be
g(x \to 0) \approx 3\sqrt{3 \pi}-\frac{\sqrt{\pi^3}}{4}x^2+{\cal O}(x^4) ~,
\ee
we can obtain the following analytic expression for the HF ground-state energy  
\be\label{eq:e_hf_exp}
\varepsilon_{\rm HF}(p,\lambda,{\tilde r})
\approx\frac{\lambda^2}{8}\left(1+p^2\right)+\frac{\pi}{24\sqrt{3}}\frac{\lambda^2}{{\tilde r}^4}\left(1-p^2\right)+{\cal O}(\lambda^4) ~,
\ee
where the second term corresponds to the sum of Hartree and exchange contributions to the ground-state energy.
According to the above expression, in a single-component system (\textit{i.e.}, $p=1$), the Hartree term is exactly canceled by the leading order contribution from exchange energy, while in an unpolarized system (\textit{i.e.}, $p=0$), only half of the Hartree energy would be canceled by exchange.
Further inspection of Eq.~(\ref{eq:e_hf}) for $p=0$ and $p=1$ reveals a quantum phase transition between unpolarized and fully-polarized phases as functions of particle density and dipole-blockade radius [see, Fig.~\ref{fig:PD}\, (b)]. This is totally equivalent to the paramagnetic-ferromagnetic phase transition in the electron liquid~\cite{GV_book}. We have also verified that the energy of a partially polarized state is always higher than either the mixed (\textit{i.e.}, unpolarized) or fully separated state. Therefore it never becomes the ground-state of a two-component system.   
The phase diagram of mixed and phase-separated states is shown in Fig.~\ref{fig:mixed-separated} in the $\lambda$-$\tilde{r}$ parameter space. 
As it is seen in the figure, for low-density and small blockade radius the system is in the phase-separated state. 
\par
In the remaining parts of this work, we will discuss the dynamical stability and collective excitations of phase-separated ground state, which can be regarded as a single-component RDL.  
\section{Density-Wave Instability of the ground state \label{sec:DWI} }
\begin{figure}
 \includegraphics[width=\linewidth]{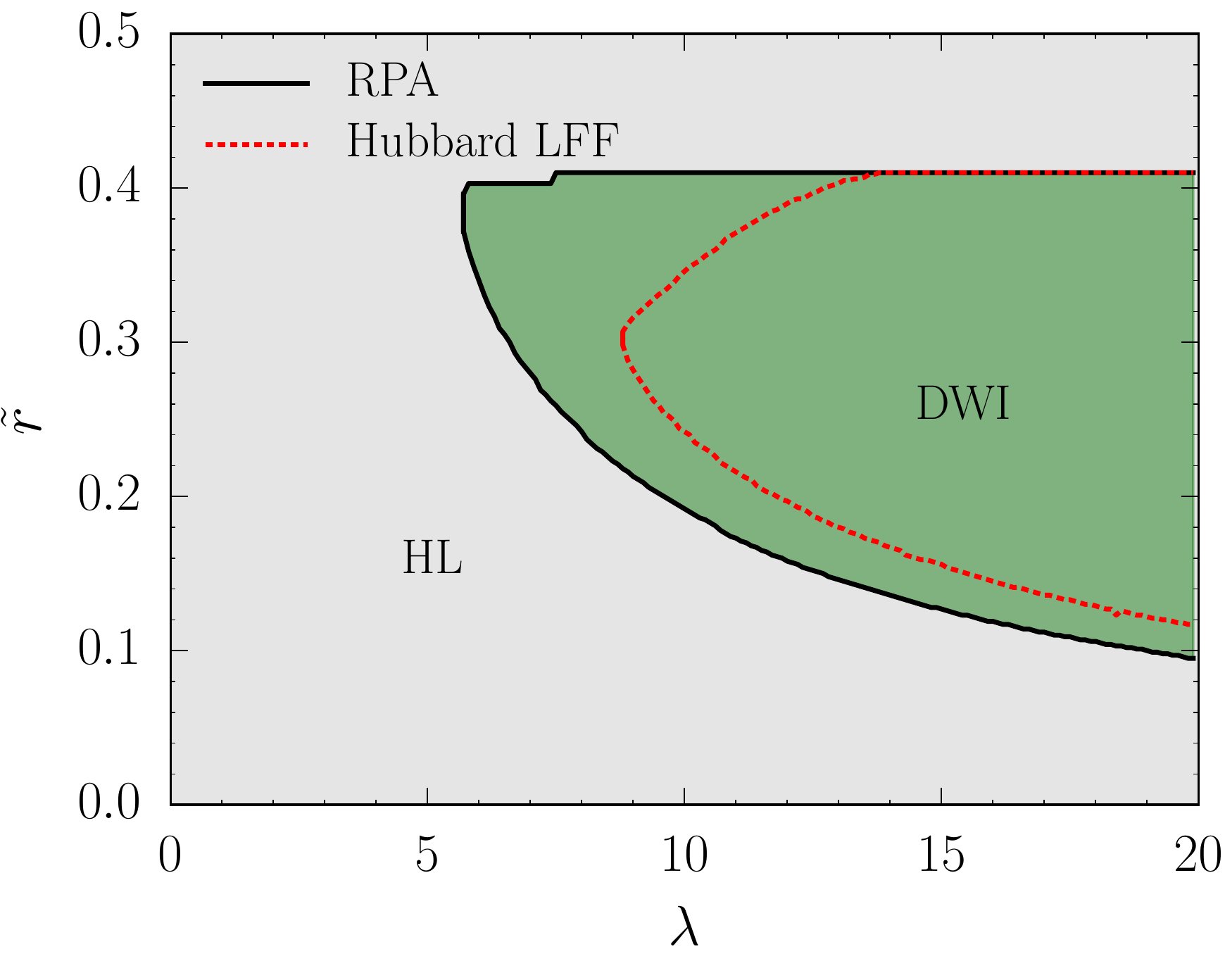} 
 \caption{The regions of stable homogenous liquid (HL) and the density-wave instability (DWI) of a single-component  two-dimensional Rydberg-dressed liquid in the $\lambda-{\tilde r}$ plane. Black solid lines show the phase boundary within the random phase approximation, and red dotted lines refer to the phase boundary obtained using the Hubbard approximation for local-field factor. 
Within both approximations, only the unstable regions with $q_I \le 2 k_{\rm F}$ for the instability wave vector have been retained. \label{fig:DW_RPA}}
\end{figure}

In this section, employing the random-phase approximation, we investigate the DWI of a single-component RDL from the poles of its static density-density response function, or equivalently from the zeros of its static dielectric function
\begin{equation} \label{eq:epsilon}
        \epsilon_{\rm RPA}(q)=1-v_{\rm R}(q)\chi_{0}(q)=0~,
\end{equation}
where $v_{\rm R}(q)$ is given in Eq.~(\ref{eq5}) and $\chi_{0}(q)$ stands for the non-interacting static density-density response function of a single-component 2D Fermi gas with parabolic energy dispersion~\cite{GV_book} 
 \begin{equation}\label{eq:chi0}
        \chi_{0}(q)=-\nu_0\left[1-\Theta (q-2 k_{\rm F})\frac{\sqrt{q^2-4 k_{\rm F}^2}}{q}\right]~.
 \end{equation}
Here $\nu_0=m/(2\pi \hbar^{2})$ is the density-of-states. 
For a given set of system parameters \textit{i.e.}, $\lambda$ and $r_c/r_0$, if the static dielectric function becomes zero at a specific wave vector $q_I$, the homogenous system becomes unstable towards a density ordered phase with wavelength $\lambda_I=2\pi/q_I$. 
It should be noted that as this instability criterion corresponds to a second-order phase transition from a homogeneous into an inhomogeneous state, a first-order phase transition of the homogenous state might precede the instability, in which case no instability would emerge
~\cite{Neilson93,Szymanski94,Swierkowski96}.
Note that as the static response function~(\ref{eq:chi0}) is negative, the density-wave instability would be feasible only in the negative regions of the interparticle interaction \textit{i.e.}, $q_I\approx 5/r_c$ within the RPA. 
In other words, the magnitude of the instability wave vector is determined by the blockade radius.
If this wave vector becomes larger than $\sim 2 k_{\rm F}$, the wavelength of the expected density ordered phase will be smaller than the average inter-particle separation. Therefore, in order to determine DWI, we should search for the solutions of Eq.~(\ref{eq:epsilon}) with $q \leq 2 k_{\rm F}$.   The phase boundary between stable homogeneous and the density-wave phases obtained within the RPA, and with the condition $q_I \leq 2 k_{\rm F}$ is illustrated in Fig.~\ref{fig:DW_RPA}. 

In the RPA, the effects of exchange and correlation are completely absent. Improvements over RPA could be achieved by replacing the bare interaction $v_{\rm R}(q)$ in Eq.~(\ref{eq:epsilon}) with an effective one
\be\label{eq:veff}
v_{\rm eff}(q)=v_{\rm R}(q)\left[1-G(q)\right]~,
\ee
where $G(q)$ is the local-field factor~\cite{GV_book}. Using the Hubbard approximation for the LFF $G_{\rm H}( q)=v_{\rm R}(\sqrt{k_{\rm F}^{2}+q^{2}})/v_{\rm R}(q)$, we have reexamined the DWI, and the corresponding result is shown by dashed lines (red) in Fig.~\ref{fig:DW_RPA}. Evidently, the homogenous liquid phase becomes more stable when the effects of exchange hole are included through the Hubbard LFF.
%

\section{Zero-sound modes\label{sec:ZSM}}
The collective density oscillations of an interacting system could be obtained from the zeros of its dynamic dielectric function. In the RPA, one should look for the solutions of
\begin{equation} \label{eq:epsilon_dyn}
1 - v_{\rm R}(q)  \Re e \,\chi_{0}(q,\omega)=0~,
\end{equation}
outside the particle-hole  continuum (PHC), \textit{i.e.}, in the regions of frequency-wave vector plane where the imaginary part of the non-interacting density response function vanishes $\Im m  \, \chi_{0}({\bm q},\omega) = 0$.

Using the analytic forms of the real and imaginary parts of the non-interacting density response function of a two-dimensional Fermi system $\chi_0(q,\omega)$~\cite{GV_book}, it turns out that it is possible to find an analytic solution for Eq.~(\ref{eq:epsilon_dyn}) outside the PHC, which gives the full dispersion of the zero-sound mode as
\be\label{eq:w_q}
\begin{split}
\omega_{\rm ZS}(q)=&v_{\rm F} q \left[1+\frac{1}{\nu_0 v_{\rm R}(q)} \right]\\
&\times \sqrt{\left(\frac{q}{2 k_{\rm F}}\right)^2+\frac{\nu^2_0 v^2_{\rm R}(q)}{1+2 \nu_0 v_{\rm R}(q)}}~.
\end{split}
\ee
\begin{figure}
 \begin{overpic}[width=\linewidth]{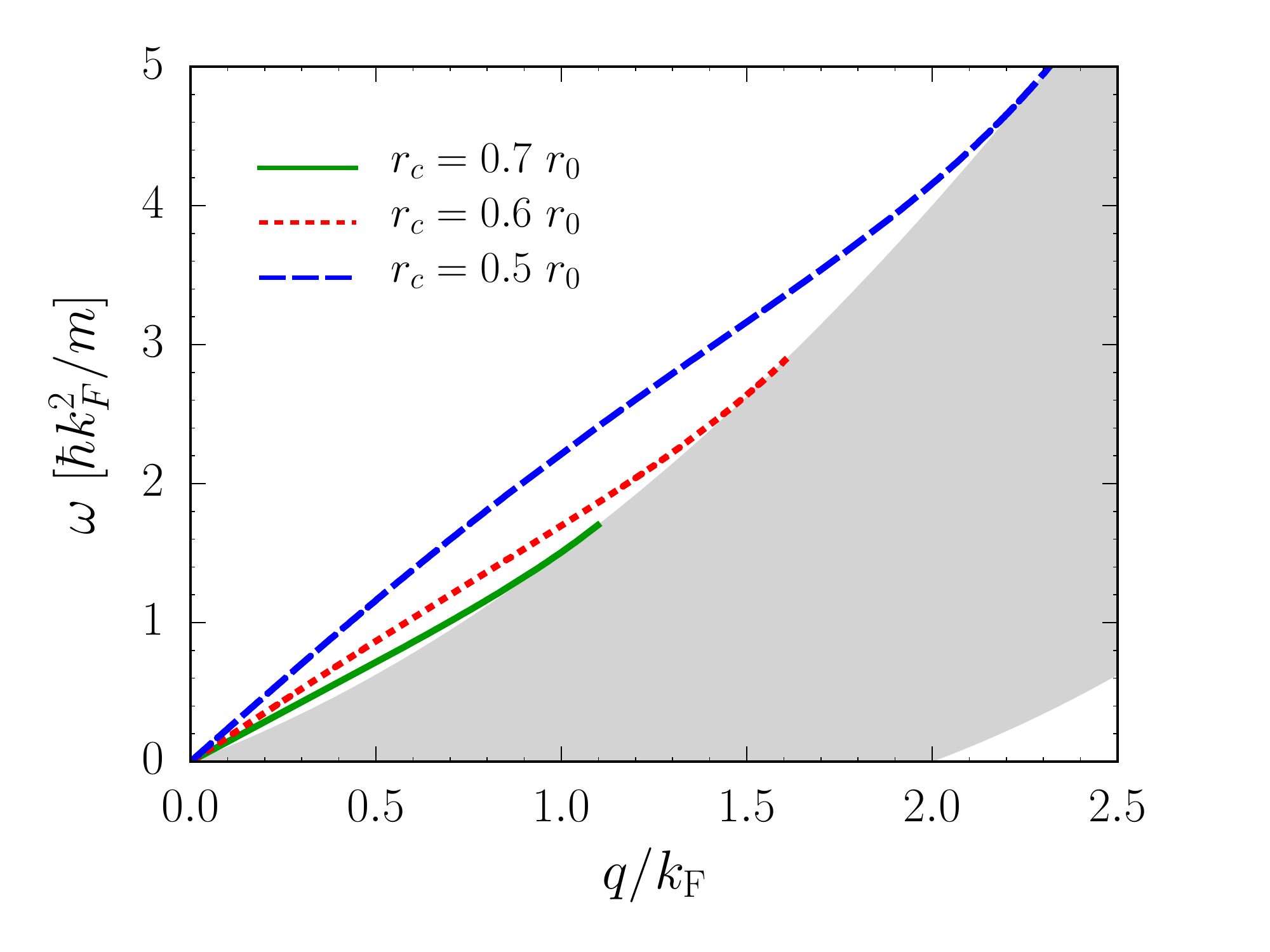}\put(-1,75){(a)}\end{overpic}
 \begin{overpic}[width=\linewidth]{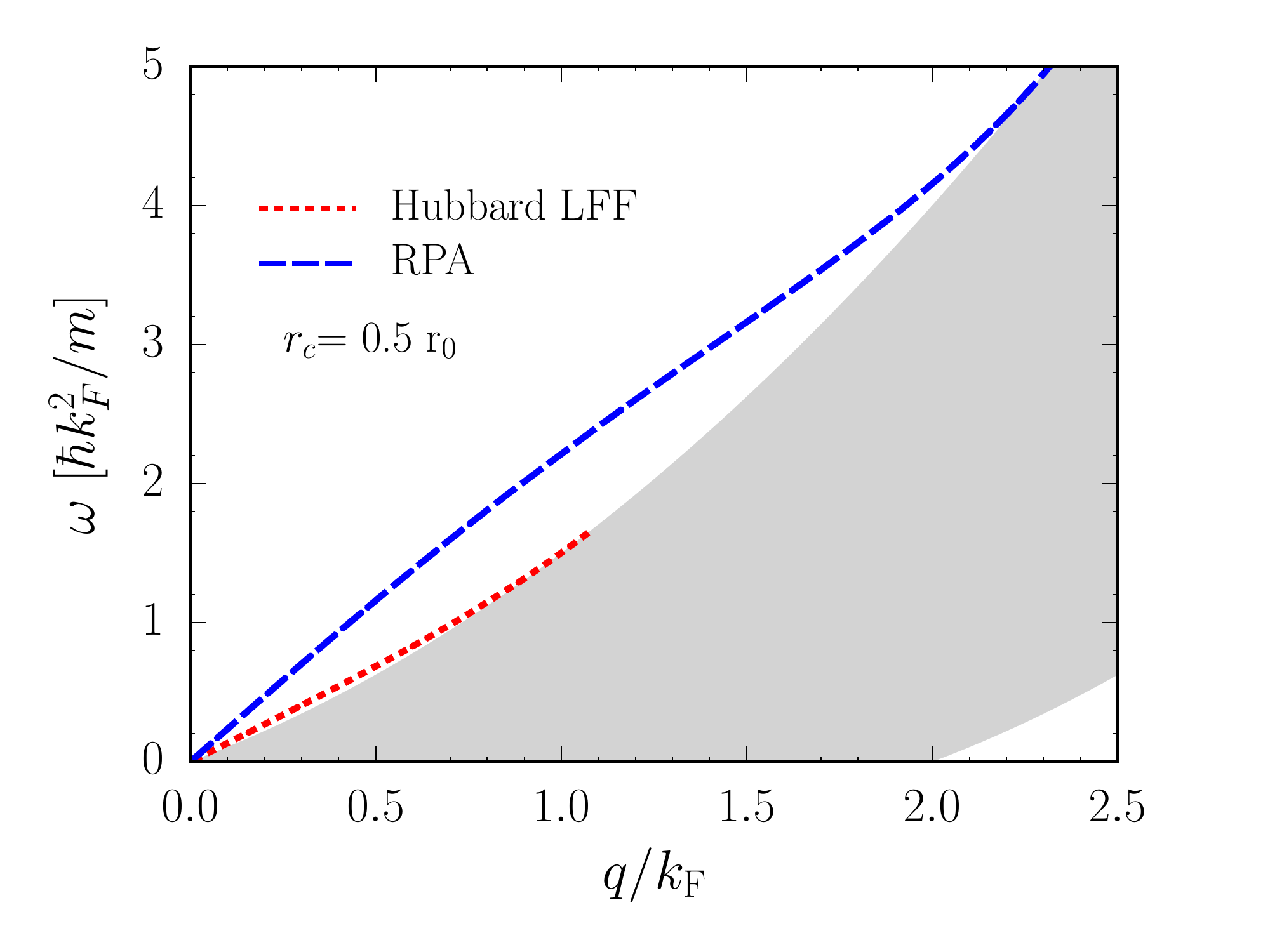}\put(-1,75){(b)}\end{overpic}
 \caption{(a) Dispersion of the zero-sound mode of a single component 2D Rydberg-dressed liquid within the RPA (in units of $\hbar k_{\rm F}^2/m$) versus $q/k_{\rm F}$ for several values of the blockade radius $r_c/r_0$. (b) Effects of the Hubbard LFF on the dispersion of zero-sound mode for $\lambda=2$. Filled areas in grey color indicate the particle-hole continuum. \label{fig:zsm}}
\end{figure}
In long wavelength limit, we have $v_{\rm R}(q \to 0) \approx u_0= {\pi^{2} D}/({3\sqrt{3}r_c^4 })$ and by plugging it in Eq.~(\ref{eq:w_q}), we find   
$\omega_{\rm ZS}( q\to 0) \approx v_{\rm ZS}\, q$ where
\begin{equation}\label{eq30}
     v_{\rm ZS} = \frac{1+\nu_0 u_0 }{\sqrt{1+2 \nu_0 u_0}} v_{\rm F} ~,
\end{equation}
is the sound velocity, with $v_{\rm F} = \hbar k_{\rm F}/m$ being the Fermi velocity.
Evidently, this velocity is larger than $v_{\rm F}$, therefore the zero-sound wave is undamped at long wavelengths. 
Effects of exchange-correlation hole in the dispersion of zero-sound mode could be obtained by replacing the bare interaction $v_{\rm R}(q)$ in Eq.~(\ref{eq:w_q}) with the effective interaction of Eq.~(\ref{eq:veff}). 
In Fig.~\ref{fig:zsm}\,(a) the full dispersion of the zero-sound mode within the RPA is plotted for different values of the blockade radius $r_c$. In Fig.~\ref{fig:zsm}\,(b) the effects of Hubbard LFF on the dispersion of zero-sound mode is depicted. Notice that exchange hole substantially suppresses the zero-sound velocity.
\par
\begin{figure}
 \begin{overpic}[width=\linewidth]{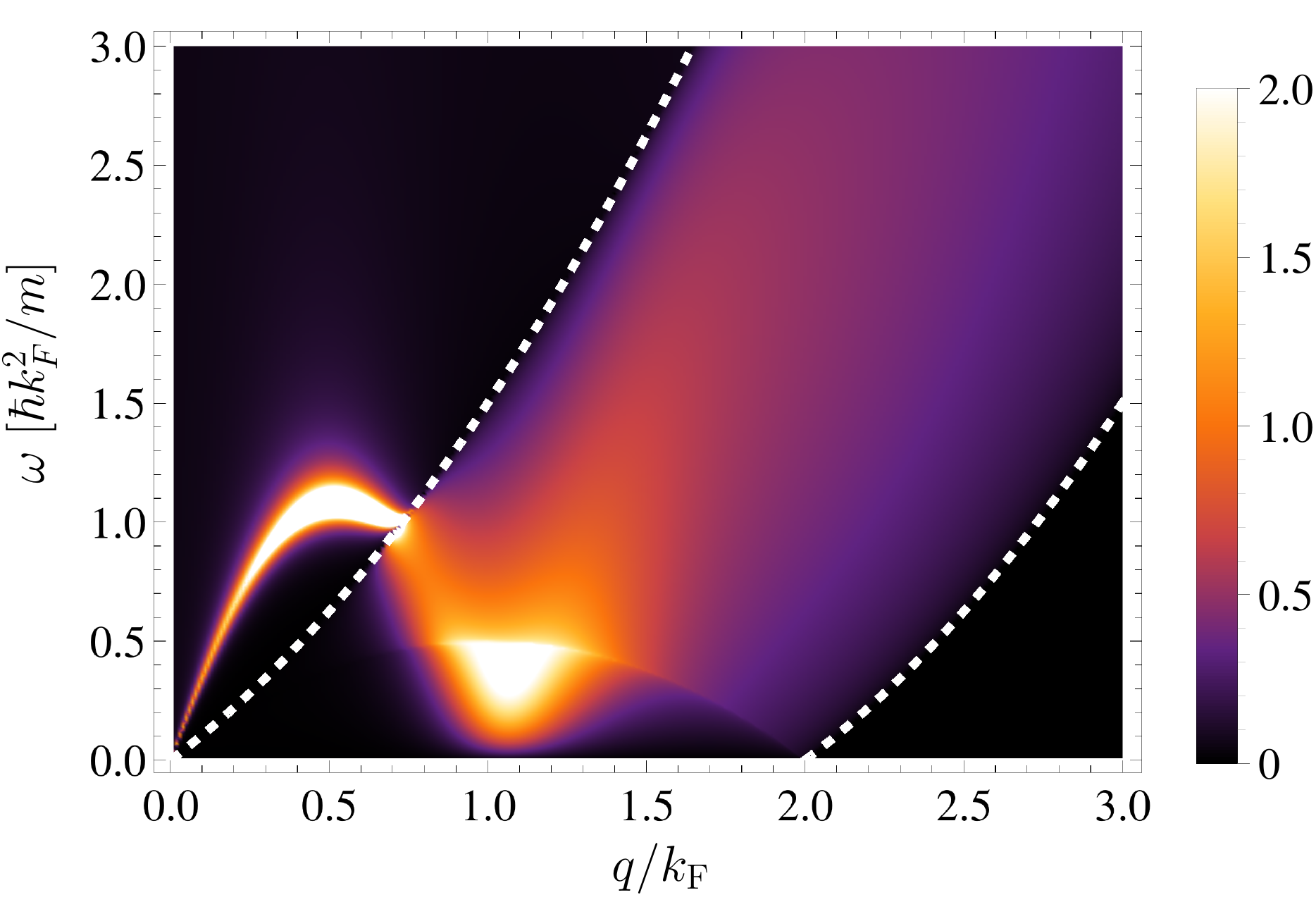}\put(-1,63){(a)}\end{overpic}
 \begin{overpic}[width=\linewidth]{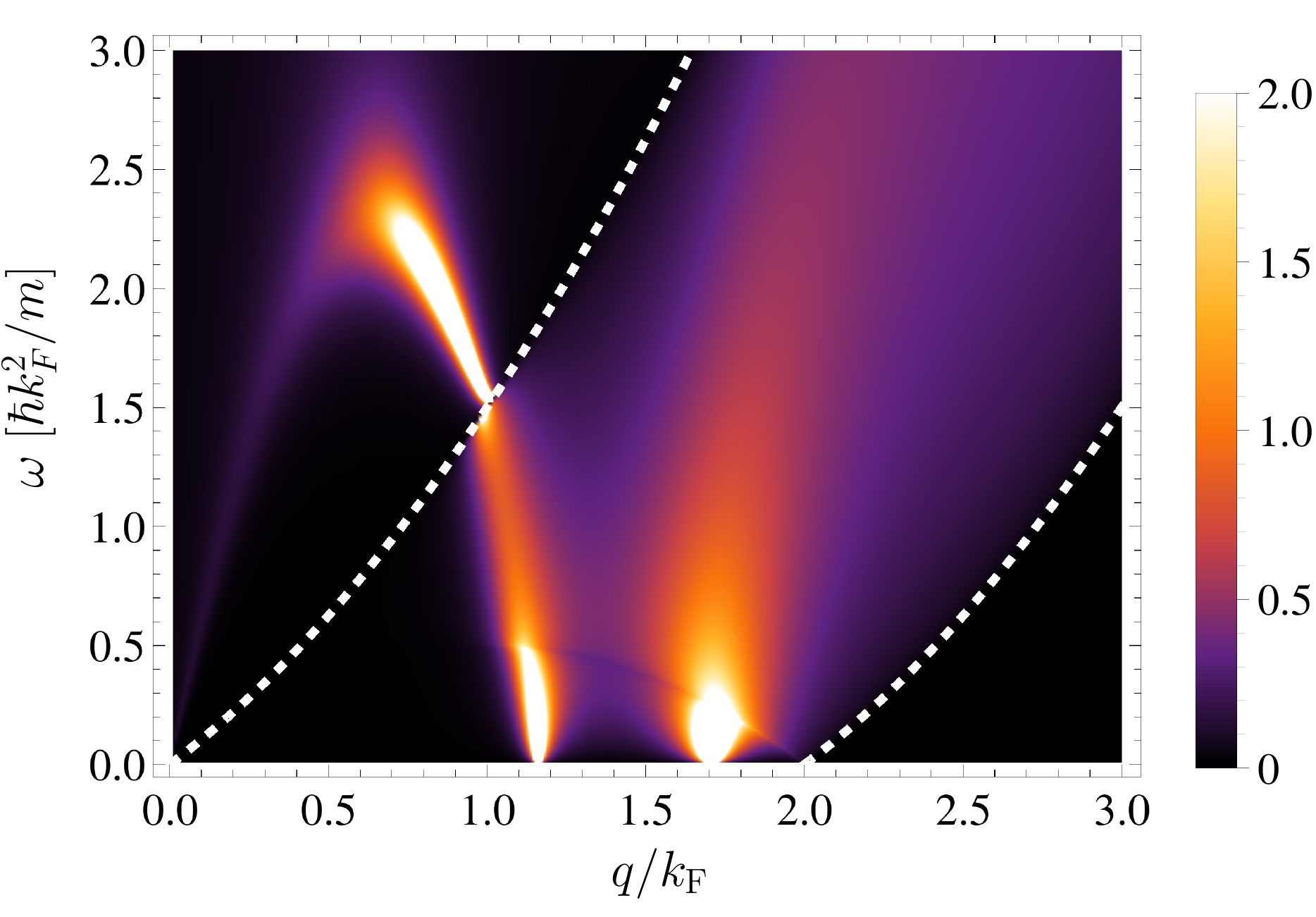}\put(-1,64){(b)}\end{overpic}
 \caption{Imaginary part of the density response function within the RPA versus $\omega$ and $q$ at a fixed density parameter $\lambda =10$, and for $\tilde{r}=0.45$ (a), and $\tilde{r}=0.3$ (b). These two parameter values for ${\tilde r}$ respectively correspond to the homogenous liquid and density-wave unstable regions of the phase diagram.
 Note that the Dirac delta peak of the spectral weight outside PHC has been broadened by $\sim 0.01$ for a better visibility. 
 \label{fig:spectral_weight}}
\end{figure}

We also investigate the oscillation strength, or the spectral weight of the collective mode, from the imaginary part of the RPA density response function
\be
\Im m\, \chi^{\rm RPA}(q,\omega)=\frac{\Im m\, \chi_0(q,\omega)}{\left|1-v_{\rm R}(q) \chi_0(q,\omega)\right|^2}~.
\ee
This spectral weight has a Dirac delta peak over the collective mode dispersion outside PHC and has contributions from both single particle and collective excitations inside the continuum. 
In Fig.~\ref{fig:spectral_weight}, the spectral weight has been depicted versus frequency and wave vector. 
For the density and blockade radius values, where the DWI is expected, mode softening and a \textit{roton}-like minimum, which reaches zero at $q\sim q_c$ is evident in the spectral weight (see, the lower panel of Fig.~\ref{fig:spectral_weight}).  
%

\section{Summary and Conclusions}\label{sec:conclusion}
In this paper, we have studied the ground-state phases of a two-dimensional Rydberg-dressed Fermi liquid. 
The physics of this system is governed by the competition between three natural length scales of the system, namely the blockade radius $r_c$, the range of van der Walls interaction $r_0$, and the average distance between atoms $1/\sqrt{n}$. In fact, the Rydberg-dressed interaction has its strongest repulsive effects for $r_c \lesssim r \lesssim r_0$.  This well explains our Hartree-Fock prediction for the phase separation in the two-component system at small blockade radius and at low densities, where the repulsive interaction of each particle with its nearest neighbors is dominant. At larger blockade radii or higher densities, the nearest neighbors of each particle will fall within its blockade radius and will be affected only by a very weak repulsive force due to the soft core of interaction.

In a single-component system, we have looked for the signatures of the instability of a homogeneous system towards density ordered phases from the singularities of its static dielectric function. The DWI is expected for high densities and at intermediate blockade radiuses. At large blockade radii, the system is effectively non-interacting, while at intermediate values of $r_c$, particles would energetically prefer to aggregate together within the blockade radius and increase the average distance between clusters of particles. Therefore inhomogeneous density phases such as density waves or quantum droplets~\cite{cinti_prl2010} would be naturally expected in this regime. 
Here, we should mention that the density-wave instability within different approximations has been also predicted for ultracold dipolar systems with anisotropic dipole-dipole interaction~\cite{yamaguchi_pra2010, sun_prb2010, parish_prl2012, vanzyl_pra2015} as well as in layered dipolar structures~\cite{emre_arxiv,block_njp2012,marchetti_prb2013}. However, its emergence in a single-layer system, with a purely isotropic dipole-dipole interaction has been the subject of much dispute~\cite{saeed_aop2014,matveeva_prl2012}. Here, we have shown that such an instability could be anticipated in a single layer of Rydberg-dressed atoms with isotropic interaction. Also, we have examined that this instability is not an artifact of the random phase approximation, and it survives when the effects of exchange-hole have been taken into account.

We have also obtained the full analytic dispersion of collective density oscillations in a single-component Rydberg-dressed Fermi liquid. This dispersion is undamped and linear at long wavelengths. Similar to the three-dimensional Rydberg-dressed liquids, mode softening in the vicinity of density instability has been observed in the spectral weight of the collective mode.  

\acknowledgments
We are grateful to Habib Rostami for useful comments and suggestions. 
This work is supported in part by Turkish Academy of Sciences (TUBA).


\end{document}